# Machine Learning for Achieving Bose-Einstein Condensation of Thulium Atoms

E. T. Davletov[1, 2], V. V. Tsyganok[1, 2, 5], V. A. Khlebnikov[1], D. A. Pershin[1, 3], D. V. Shaykin[1, 2] and A.V. Akimov[1,3,4]

[1]Russian Quantum Center, Business Center "Ural", 100A Novaya Str., Skolkovo, Moscow, 143025, Russia

[2]Moscow Institute of Physics and Technology, Institutskii per. 9, Dolgoprudny, Moscow Region 141701, Russia

[3]PN Lebedev Institute RAS, Leninsky Prospekt 53, Moscow, 119991, Russia

[4]Texas A&M University, 4242 TAMU, College Station, Texas, 77843, USA

[5] National University of Science and Technology, Leninsky Prospekt 4, Moscow, 119049, Russia, email: akimov@physics.tamu.edu

Bose-Einstein condensation (BEC) is a powerful tool for a wide range of research activities, a large fraction of which are related to quantum simulations[1–5]. Various problems may benefit from different atomic species, but cooling down novel species interesting for quantum simulations to BEC temperatures requires a substantial amount of optimization and is usually considered as a hard experimental task[6–12]. In this work, we implemented the Bayesian machine learning technique[13] to optimize the evaporative cooling of thulium atoms and achieved BEC in an optical dipole trap operating near 532 nm. The developed approach could be used to cool down other novel atomic species to quantum degeneracy without additional studies of their properties.

For a number of problems in quantum simulations, long-range interactions are of great interest[2]. To address this demand, lanthanide atoms such as erbium and dysprosium were cooled down to quantum degeneracy and successfully used for quantum simulations[3–5]. Thulium also belongs to the lanthanide series of chemical elements, but compared to previously cooled Er and Dy, it has only one hole in the f-shell. This leads to a slightly smaller dipole moment in the ground state of $4\mu_B$, but thulium has a simpler level structure and a less dense Fano-Feshbach resonances spectrum, thus simplifying the control of the interactions for this atom[14]. The lower dipole moment could be overcome by using thulium atoms in a 532 nm optical lattice, thus increasing the interaction strength via a shorter distance.

Reaching BEC is usually done via evaporative cooling[15]. The evaporative sequence needs to be carefully optimized to minimize the loss of atoms while maximizing the phase space density growth. Many studies have been devoted to the elaboration of a better recipe for the evaporation sequence optimization process[16–18]. Most of these techniques require accurate characterization of the trap geometry and loss mechanisms; they also use several simplifications, such as a high and fixed truncation parameter $(\eta)$ or adiabaticity, which are not always satisfied. Furthermore, these recipes usually do not consider difficulties and/or opportunities that arise in specific cases such as dynamically shaped traps and the existence of dipolar interactions. Hence, most groups use a simple time-consuming stepwise optimization procedure, with a proposed sequential adjustment of parameters controlling the trapping potential at each time step[6–12].

The initial cooling of thulium atoms was realized in a magneto-optical trap operating on a narrow-line transition with a wavelength of approximately 530.7 nm and Zeeman slowing at a strong 410 nm transition[19]. An optical dipole trap is realized using a 532 nm laser[20]. The narrow transition of approximately 530.7 nm that is only 345 kHz wide could cause some minor resonance scattering and additionally complicate the evaporative cooling[21]. The following difficulties complicate the search for a good evaporative sequence. 1) Thulium has a short lifetime in the dipole trap[21] (2.8 s at the optical dipole trap at temperatures just below the magneto-optical trap) compared to other lanthanide species ( >7 s for Er[6], > 21 s for Dy[7], and >140 s Cr[8]. 2) The maximum depth of the dipole trap ~110 uK is only 4.5 to 5.5 times higher than the temperature of the cloud (15 for Dy[7] and 13 for Er[6]). 3) Imperfections in trap characterization especially due to time-dependent trap geometry and the complex dependence of atom polarizability on the optical dipole trap (ODT) beam geometry and light polarization[21]. 4) The ODT has a considerable atom loss rate, but there is not a complete understanding of the loss mechanisms at the moment. 5) There are generally no models that are able to capture the process of transferring atoms from a single beam ODT to a crossed beam ODT, and this process strongly influences the evaporation efficiency.

6) In addition, the intermediate stages of the cooling transition from a single ODT to a crossed ODT create additional ambiguity in the number of atoms due to the presence of hot atoms inside one of the traps that are not captured in the crossed ODT region.

This list makes the building of an evaporative process model almost impossible and demands a substantial amount of additional work to characterize the trap and atom properties. The experimental optimization of many mutually related parameters is quite difficult on its own.

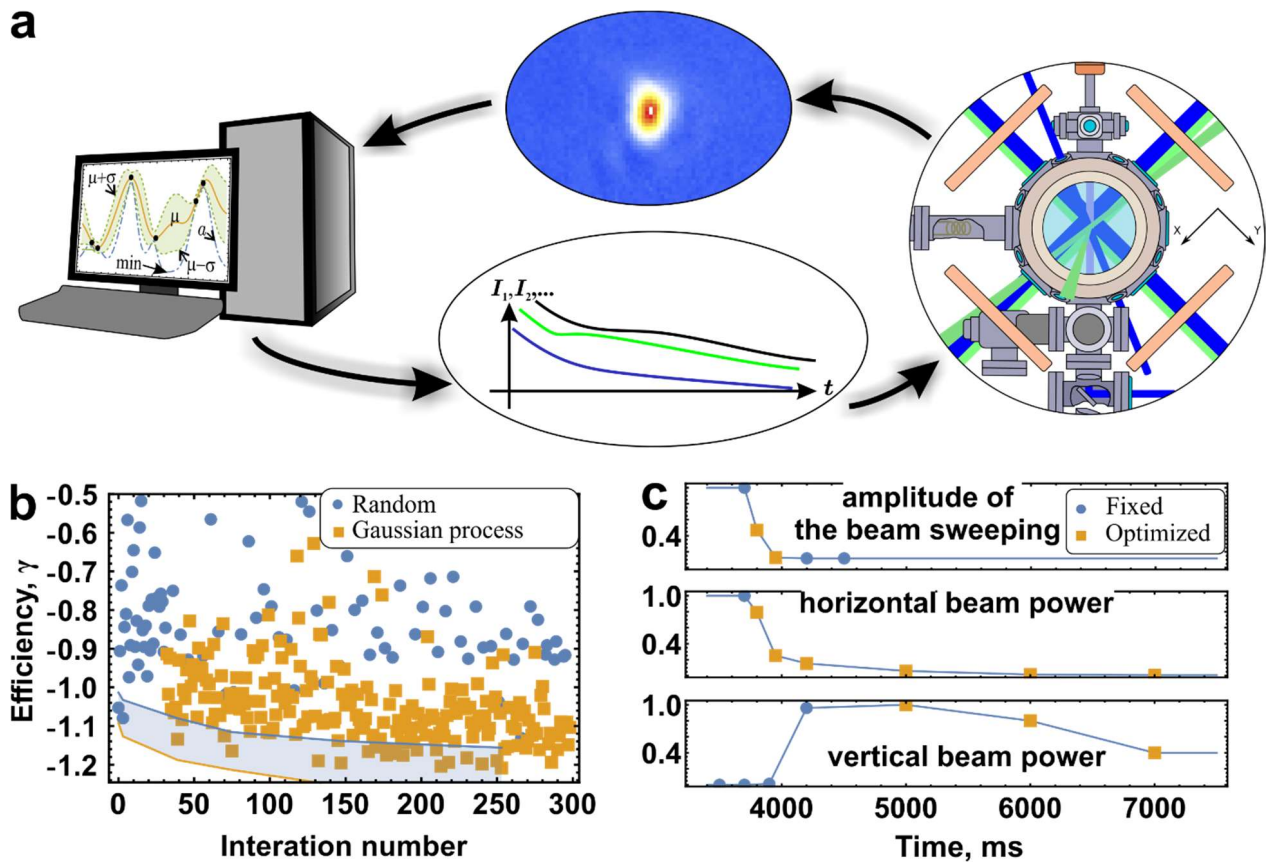

*Figure 1 a – Idea of the optimization. The parameters to regulate are the parameters of evaporation cooling –the power of the beams and the amplitude of beam sweeping. b – Example of the optimization algorithm result. In this run, the $\gamma$ parameter was optimized. The blue dots are generated randomly (the first 30 points seed the GP model, and each additional 5 are to sample unbiased data). The orange dots are iterations of the algorithm. The green area is the confidence interval ($\sigma_{conf}$ wide) for the points minimizing the GP model $\mu_n(x)$ at certain iterations. c – Final evaporation sequence. The yellow dots are the parameters learned by the algorithm. The horizontal beam intensity and the sweeping and vertical beam intensity are linearly interpolated.*

We note that while some of these problems (4, 6, and partly 3) could be resolved through additional efforts, there is generally no need for those efforts from the viewpoint of evaporative sequence optimization as long as there is a sequence that provides efficient BEC formation. Instead, to overcome these difficulties, we use an adaptive experimental design based on a Bayesian machine learning technique. Adaptive Bayesian optimization is a well-known statistical approach for optimization of expensive-to-evaluate functions[22]. It is widely used to tune model hyperparameters in machine learning algorithms[23] and to solve some hard optimization problems in physics[24–28]. The idea of this approach is to avoid the direct optimization of the real system and instead build a surrogate model of the system at each step and utilize it to wisely choose the next point to sample from the real system[29].

In more detail, we use the most common class of surrogate models, the Gaussian Process (GP) models[13]. The GP model outcome for experiment $f(\vec{x})$ (the evaporation efficiency $\gamma$ in our case, see below) started with the vector of the controlling parameters $\vec{x}$ as a random variable. At any given vector $\vec{x}$, this model assumes that the distribution $p(f(\vec{x}))$ is a normal distribution $N(\mu(\vec{x}), \sigma(\vec{x}))$ with a mean $\mu(\vec{x})$ (see the inset in Figure 1a for the case with 1D $\vec{x}$) and a standard deviation $\sigma(\vec{x})$ (Figure 1a). These parameters are calculated at each step (see Supplementary Information) using already observed experimental results. With the GP model, one can choose a new vector of parameters $\vec{x}^*$, which is chosen such that it minimizes the function $a(\vec{x}^*) = \mu(\vec{x}^*) + \alpha \sigma(\vec{x}^*)$ (Figure 1a). The $a(\vec{x}^*)$ is usually called the upper confidence bound (UCB)[30] acquisition function, which has some scalar fixed parameter $\alpha$ (see Supplementary Information). Finally, the optimization process implemented using software packages[26,31] can be represented as follows:

1. The algorithm measures $n$ outcomes of the experiment $[f(\vec{x}_1), ..., f(\vec{x}_n)]$ at some randomly chosen points $[\vec{x}_1, ..., \vec{x}_n]$ and uses them to build (seed) the initial GP model (see Figure 1b, blue points).

2. The GP model algorithm calculates the acquisition function $a(\vec{x})$ and searches for the $\vec{x}^*$ that minimizes its value $\vec{x}^* = \arg\min(a(\vec{x}))$ (see Figure 1c).

3. The experiment starts at the new value of the control parameters $\vec{x}^*$, $f(\vec{x}^*)$ is measured (see Figure 1b, orange points), and the GP model is rebuilt (see Supplementary Information).

4. Steps 2-3 repeat until the desired number of iterations is reached. Typically, the number of iterations was chosen large in our experiments, and the code was stopped when no improvement could be seen in the optimization criteria.

To perform optimization, one needs to choose the optimization criteria $f(\vec{x})$ and define the optimization parameters $\vec{x}$. The larger the number of controlled parameters, the closer the algorithm can converge to the global minimum of the problem. The convergence time also increases with the number of parameters, thus forcing the researcher to limit possible "knobs". In our case, after the pre-cooling stages in the magneto-optical trap (MOT) and the atoms are spin-polarized, they are loaded into the horizontal dipole trap, the aspect ratio of which is dynamically shaped by an acousto-optical modulator (AOM) to increase the efficiency of loading atoms from the MOT to the ODT[20] (see Methods). After the MOT is switched off, one needs to first efficiently turn off the beam scanning to increase the atomic density; then, during the evaporative cooling, the second beam (the vertical beam in our case) must be turned on to provide a better localization of atoms and a higher density in the trap. Finally, evaporation cooling is performed until BEC is achieved. For the entire procedure, the possible degrees of freedoms are the shape of the ramp for the amplitude of the beam scanning, the shape of the ramp for the intensity of the horizontal trap beam, and the shape of the ramp for the intensity of the vertical trap beam. Each initial and final time point for the ramps are the parameters and the actual shapes.

The optimization goal $f(\vec{x})$ was set to be the efficiency of evaporation cooling $\gamma$:

$$\gamma = -\ln\frac{PSD}{PSD_0} \bigg/ \ln\frac{N}{N_0} \tag{1}$$

where $PSD_0$ is the initial phase space density, $PSD$ is the final phase space density, $N_0$ is the initial number of particles in the ODT, and $N$ is the final number of particles in the ODT.

To approximate the ramp shapes, the entire sequence was split into linear fragments. The algorithm was able to optimize the end points of each segment. Thus, we picked several specific times at which the parameters could be changed by the optimization algorithm. The end and start times of each segment were not varied, but the power of the beams and the horizontal beam waist were varied (see Figure 1c).

The performance of the optimization algorithm for a magnetic field of 4.04 G on step 1 is demonstrated in Figure 1b. The algorithm clearly converges, although there is a rather high level of noise, which is mainly due to the fit error in the estimation of the temperature and the number of atoms (see Supplementary Information for more details). By performing a search in the parameter space, the algorithm converges to a $\gamma$ value of approximately $1.6 \pm 0.1$ via the evaporation sequence, as demonstrated in Figure 1c. The best stepwise optimization that was achieved manually was $\gamma \approx 0.78$, which was not enough for BEC formation. The sequence found by the algorithm finally led to a bimodal distribution of the atomic densities, which is characteristic for the Bose-Einstein condensation (Figure 2a). Although the value of $\gamma$ achieved by the algorithm is still slightly smaller than that found for the Er and Dy experiments[6,7], it is enough to achieve BEC.

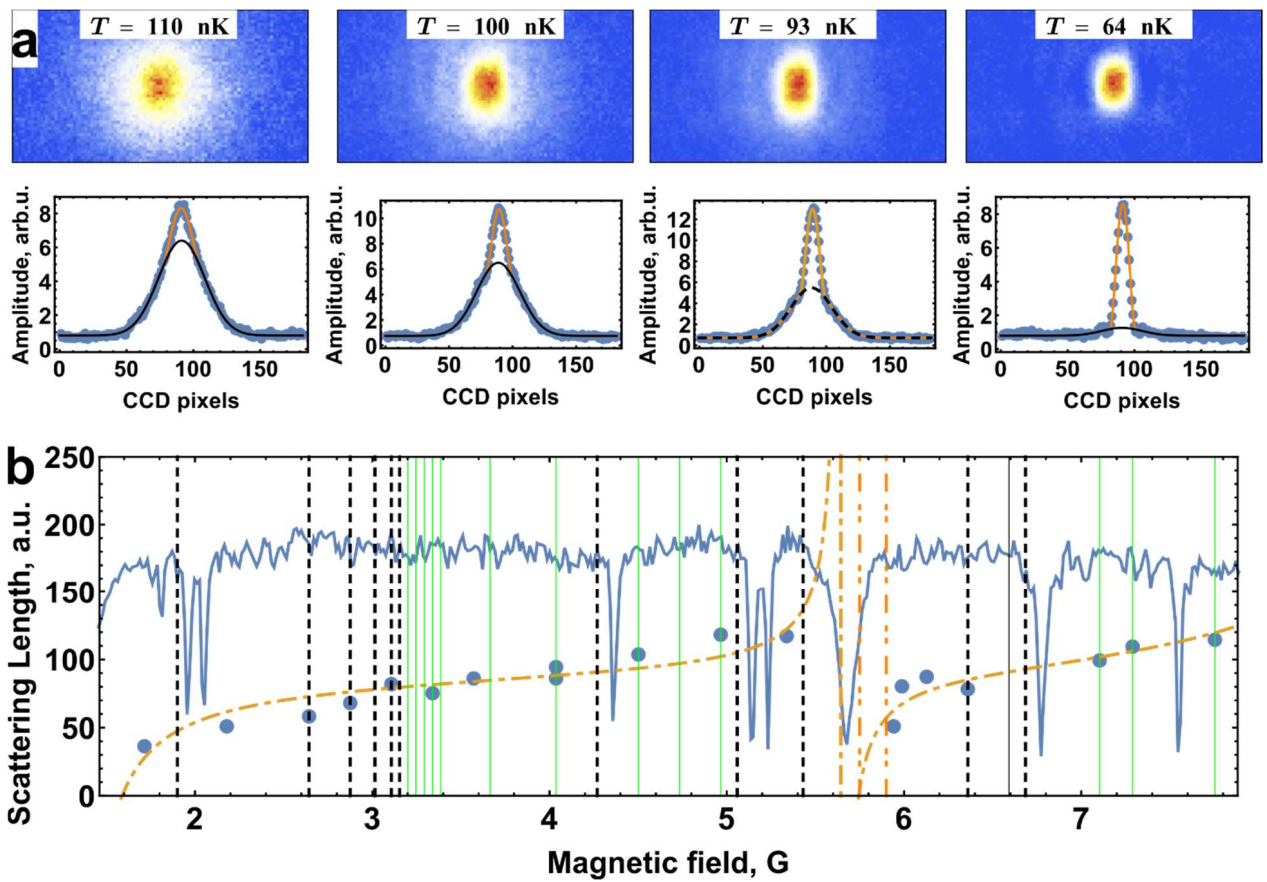

*Figure 2 a – Typical photos of the atomic cloud with BEC inside and the corresponding bimodal density distribution at different temperatures. The photos were taken at 14.5 ms of expansion. The orange solid line represents the fit by the bi-modal distribution (see Supplementary Information for details), the black dashed line shows the contribution of the thermal cloud. b – Measured scattering length in atomic units versus magnetic field (dots) and the fit of the scattering length (orange dash-dot line), taking into account only broad Feshbach resonances at 1.34, 5.68 and 9.53 G. The blue solid line represents the loss spectrum*

We note that while this optimization does not require any prior knowledge of the collisional properties of thulium, the optimization results will obviously depend on them. Figure 2b demonstrates the availability of BEC with the cooling sequence, as found by the algorithm at a fixed magnetic field $B = 4.04$ G and at different magnetic fields, along with the measured scattering length (see Methods). One can see that outside of the sharp resonances, the scattering length continuously changes due to the presence of broad Feshbach resonances. Although the variation in the scattering length strongly affects the formation of BEC, there are rather wide ranges of positive scattering lengths in which the sequence found does lead to BEC formation. Thus, the identified sequence is quite stable with respect to changes in the strength of the interatomic interactions and potentially some associated losses.

The bimodal distribution[32] depicted in Figure 2a shows a striking manifestation of BEC formation. To demonstrate the coherence properties of our atomic cloud, we performed additional experiments. One of the manifestations of coherence in the atomic cloud is its asymmetric expansion, which is even more asymmetric in the presence of dipole-dipole interations[33]. As a coherent state, BEC should "diffract", i.e., the more constrained direction of the cloud should expand more rapidly. Indeed, this behavior could be observed in our cloud, as demonstrated in Figure 3a. At small expansion times, our imaging system is unfortunately not able to correctly measure the cloud size due to diffraction on the atomic cloud and resolution limitations, but as the cloud expands, the image starts to follow the predicted behavior (see Supplementary Information). In contrast, the thermal cloud tends to expand more and more symmetrically with time, with the aspect ratio of the cloud sizes asymptotically approaching 1 because wave interference does not take place in this case. Indeed, just above the critical temperature, we observe uniform expansion of the atomic cloud (see Figure 3a).

Nevertheless, a similar distribution and an asymmetric expansion could appear in the hydrodynamic regime[34]. This regime may occur if the Knudsen criterion, $Kn \ll 1$, is satisfied (see Methods). We found that $Kn > 1$ for each axis in all stages of the thermal cloud to the formation of BEC (see Supplementary

Information). In addition, at the hydrodynamic limit, the velocity distribution of the cloud along the tight axis should have a peak over the thermal distribution while having velocities reduced beneath the thermal distribution along the weak axis. In contrast, in our experiment, both velocity distributions along the weak and tight directions have peaks (see Supplementary Information).

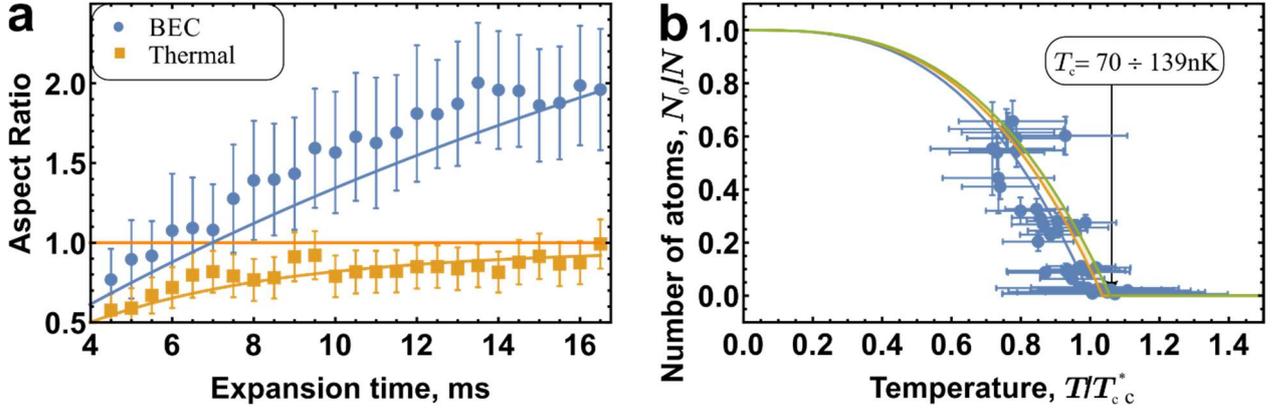

*Figure 3 a – Aspect ratio for the BECs and the visible sizes of the thermal cloud versus the time of flight after releasing from the ODT. The lines represent theoretical simulations (see Methods), and the dots represent experimental data. The red horizontal line indicates the asymptotic behavior of the thermal cloud. b – Number of condensed atoms $N_0$ normalized by the total number of atoms $N$ versus the ratio between the thermal part temperature $T$ and the critical temperature $T_c^*$ of the cloud, including interaction corrections[35]. The dots are measured values, and the solid line is the theoretical dependence, including the dipole-dipole interaction between the atoms and the finite number of atom corrections. The dashed line represents the simulations without accounting for interactions for the range of identified critical temperatures to the solid line plot (see Supplementary Information for more details).*

The bimodal distribution obtained by evaporative cooling with the slightly different initial number of atoms provides the experimental dependence of the condensate fraction on the temperature (see Figure 3b). This result is in agreement with the well-known result for the number of atoms in the condensate depending on temperature:

$$\frac{N_{BEC}}{N} = 1 - \left(\frac{T}{T_C}\right)^3, \qquad (2)$$

where $N_{BEC}$ is the number of condensed particles, $N$ is the whole number of particles and $T_C$ is the critical temperature (see Supplementary Information for more details).

In conclusion, we have demonstrated the Bose-Einstein condensation of thulium atoms using a statistically justified method (namely, GP-based Bayesian Optimization) to choose the evaporative sequence. We found that while there is no complete characterization of the experimental setup, the GP model is able to optimize the evaporative cooling to achieve BEC without much knowledge of the atomic system. Although it is less descriptive, this approach is goal-oriented and far more universal and can thus be used in a wide range of experimental tasks. More specifically, we demonstrated the formation of BEC of thulium atoms, confirmed the results with a bimodal distribution, demonstrated the dependence of the condensate fraction on the temperature, and showed the coherent behavior of the cloud. We also explored BEC formation for a wide range of background scattering lengths, demonstrating the robustness of the evaporative sequence to variations in the scattering length.

This research was supported by the Russian Science Foundation grant #18-12-00266.

## Methods

After reloading polarized atoms from the magneto-optical trap into the one-dimensional horizontal optical dipole trap in the sweeping regime, we typically have approximately $6 \cdot 10^6$ atoms with a temperature of 20 $\mu K$, which corresponds to a phase space density of approximately $2.3 \cdot 10^{-5}$. The trap frequencies were measured by the standard technique of trap-frequency measurements[21] to be in the sweeping trap $\nu_x; \nu_y; \nu_z = 5.8 \pm 0.1; 160 \pm 2; 1538 \pm 20\,\text{Hz}$. The sweeping of the ODT beam is realized using an acousto-optical modulator, which was fed by a calibrated voltage control oscillator[36]. Our evaporative cycle took place in the presence of a small magnetic field of 4.04 $G$, which is not very far from Feshbach resonance. We also measured the scattering length in this magnetic field $a_0 = 90 \pm 11 a_{Bohr}$, where $a_{Bohr} = \hbar / m_e c \alpha$ is the atomic unit for the scattering length. The typical frequencies in the BEC regime were measured to be $\nu_z = (197 \pm 1)\,\text{Hz}$; $\nu_x = (23 \pm 1)\,\text{Hz}$; and $\nu_y = (137 \pm 6)\,\text{Hz}$.

To calculate the theoretical curve in Figure 3a, we used equations from[33] but changed the sign before the terms containing derivatives of the function $f$, as there appeared to be a typo in the original paper.

The main uncertainties in the measurements of the atomic numbers and temperatures are as follows: an uncertainty in the number of atoms of 10% (see Supplementary Information); errors in the temperature of approximately 6% and 10% in the $z$- and $x$-directions, respectively; and errors in the trap frequencies estimated[21] as 1% and 4% for the $z$- and $x$-directions, respectively.

The Knudsen number is defined as $Kn = \omega_i \tau_c$. Here, $\omega_i$ with $(i \in \{x,y,z\})$ is the trap frequency, and $\tau_c = \sqrt{2} n_0 \overline{v_{th}} \sigma$ is the average time between atomic collisions in the trap with a peak density $n_0$, a thermal velocity $\overline{v_{th}}$ and an elastic-scattering cross section $\sigma$.